\documentclass{appolb}
\usepackage{epsfig}
\newcommand{\asb}{{\alpha}_s}
\newcommand{\be}{\begin{equation}}
\newcommand{\ee}{\end{equation}}
\newcommand{\bea}{\begin{eqnarray}}
\newcommand{\eea}{\end{eqnarray}}
\newcommand{\oms}{\omega_\mathrm{s}}
\newcommand{\omp}{\omega_\mathrm{I\!P}}

\begin{document}
\title{Running coupling and BFKL pomeron%
\thanks{Presented at X International Workshop on Deep Inelastic Scattering
(DIS2002), \\ Cracow, Poland, 30 April - 4 May 2002 }%
}
\author{Anna M. Sta\'sto
\address{INFN Sezione di Firenze, Sesto Fiorentino (FI), Italy \\
H. Niewodnicza\'nski Institute of Nuclear Physics, Krak\'ow, Poland}
}
\maketitle
\begin{abstract}
We show that in the case of the BFKL pomeron with running coupling 
the diffusion pattern is strongly modified
and is characterised by the sudden tunneling transition
to the non-perturbative regime.
We suggest that by using the $b$ - expansion method one can suppress the non-perturbative Pomeron and isolate purely perturbative part of the gluon Green's function.
\end{abstract}
  
One of the major issues in the high energy limit of QCD is the problem of interplay between the perturbative and non-perturbative regimes.
The BFKL Pomeron in the leading logarithmic approximation \cite{LLBFKL}
results in the equation which exhibits the characteristic diffusion pattern \cite{BARLOT}. In the process with two, comparable hard scales (for example $\gamma^*-\gamma^*$, or forward jet/$\pi^0$ in DIS) the distribution of the transverse scales of the gluons broadens with increasing rapidity, with the width  $\Delta t \simeq \sqrt{\asb Y}$ where $t=\ln k^2/\Lambda^2$ with $k^2$ being the transverse momentum of the gluon and $\Lambda$ being a QCD parameter. 
Thus, for sufficiently high energies the distribution of the gluon momenta
will always reach non-perturbative regime. This picture of diffusion is well rooted in the case of the fixed $\asb$ coupling. 
The effect of subleading corrections \cite{NLLBFKL} is, among other, the running of the QCD coupling. 
In this case one would expect small modification of the diffusion picture, namely that the distribution of the momenta will develop an asymmetry towards lower
scales. The exponent of the cross section gains additional term $b^2 \asb^5 Y^3$ apart from the usual leading $\asb Y$ term, where $Y$ is the rapidity of the process.
This picture holds however for rapidities which are not too large.
In \cite{TUNNEL} it has been pointed out that the transition to the non-perturbative regime can occur as a sudden tunneling effect rather than a gradual diffusion process. Instead of slow increase of the width of the distribution of the momenta, one observes a sudden transition from the perturbative scales $t'\simeq t=\ln Q^2/\Lambda^2$ into the non-perturbative regime $t' \simeq \bar{t} \ge 0$, where $\bar{t}=\ln Q_0^2/\Lambda^2$ is the scale set by the regularisation of the running coupling. The characteristic feature of this effect is that it occurs without any passage through the intermediate scales\footnote{It has to be stressed that the unitarity effects can in principle
change significantly the phenomenon of tunneling. It was noticed in
\cite{GBMS} that in the case of the non-linear small $x$
evolution equation, the generation of the
saturation scale $Q_s(x)$ leads to the suppression of diffusion into
the low scales $k<Q_s(x)$ and the distribution of the momenta is
driven towards the perturbative regime.}.

To gain insight into the effect of the tunneling we shall consider the small $x$ evolution which is controlled by the BFKL equation of the form
\begin{equation}
\frac{\partial G(Y;t,t_0)}{\partial Y} = K \otimes G
\label{eq:bfkl}
\end{equation}
where $K$ is the BFKL kernel and $G(Y,t,t_0)$ gluon Green's function
evaluated at rapidity $Y=\ln 1/x$ and scales $t$ and $t_0$ with initial
condition
\mbox{$G(Y=0;t,t_0) = \delta(t-t_0)$}.
Kernel $K=\asb K_0$ is the usual BFKL kernel in the leading logarithmic approximation, but we additionally introduce the running of the coupling $\asb(t)$.

The detailed mechanism of the tunneling is illustrated in Fig.1 where we show the contour plots in the $t$ and $Y$ plane of the function
\begin{equation}
f(Y,y;t,t') \; = \; \frac{G(y;t,t')G(Y-y;t',t)}{G(Y;t,t)}
\label{eq:ffunction}
\end{equation}
which illustrates the change of distribution of the transverse momenta 
during evolution between two points: from $(0,t_0)$ to  $(Y,t)$.
At the beginning of the evolution $Y<50$ the solution exhibits typical diffusion pattern and the broadening of the distribution. Around $Y=60,70$ the tunneling transition occurs in which emerges second region concentrated around  non-perturbative scale $\bar{t}$. At higher rapidities the  evolution is concentrated only in the non-perturbative region. 
One can estimate the value of rapidity at which the tunneling transition takes place. We just have to compare the contribution to the Green's function which comes from the tunneling configuration with that coming from the perturbative evolution.
The tunneling configuration can be written as:
\begin{equation}
G_{\rm tunnel}(Y;t,t) \sim e^{-(t-\bar{t})/2} e^{\omp Y}  e^{-(t-\bar{t})/2}
\end{equation}
where the first term corresponds to branching from $t$ to $\bar{t}$, second to the evolution at $\bar{t}$ with a non-perturbative Pomeron intercept  $\omp$ and the last term corresponds to branching from $\bar{t}$ back to $t$. Instead the configuration  which comes from the normal perturbative evolution is just
\begin{equation}
G_{\rm pert}(Y;t,t) \sim e^{\oms(t) Y}
\end{equation}
where $\oms(t)$ is the effective perturbative saddle point exponent.

\begin{figure}[htbp]
  \begin{center}
    \epsfig{file=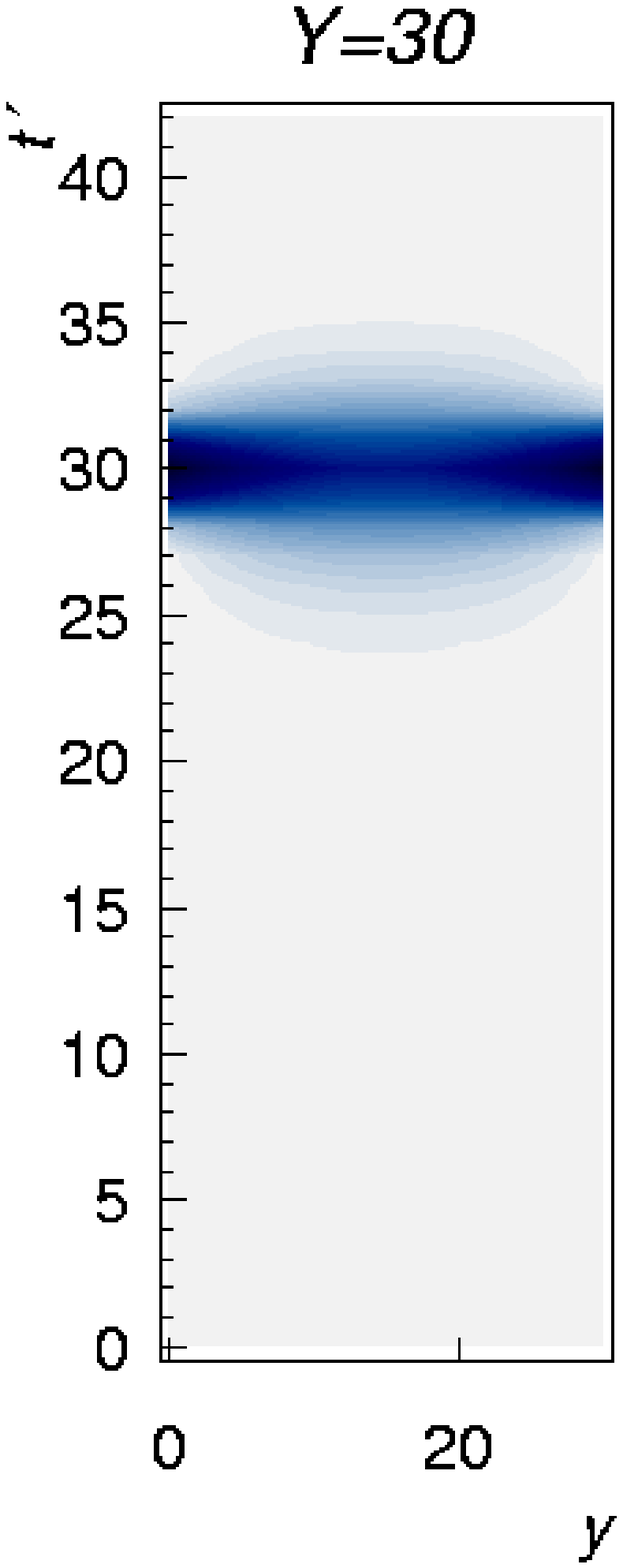,height=0.22\textheight}
    \epsfig{file=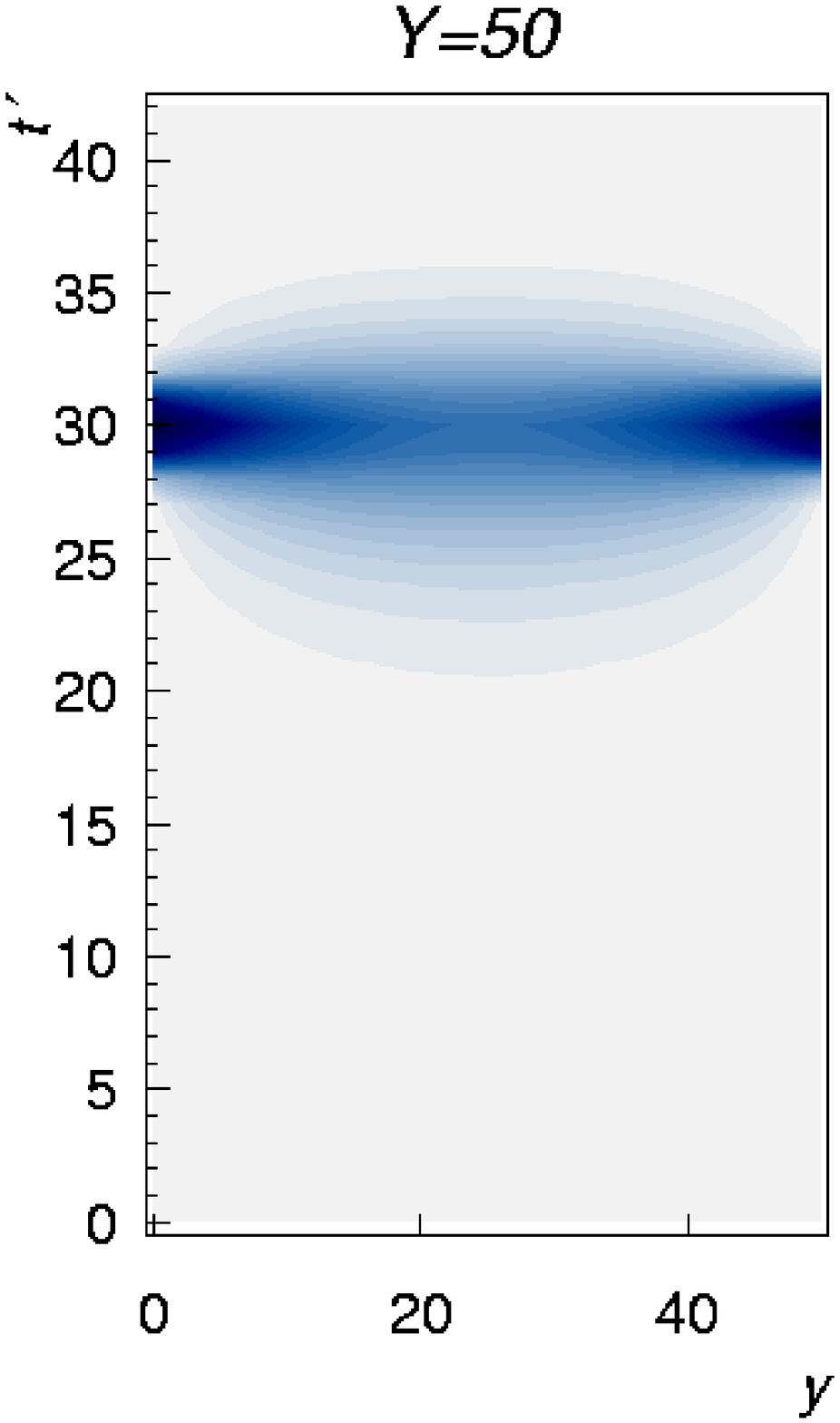,height=0.22\textheight}
    \epsfig{file=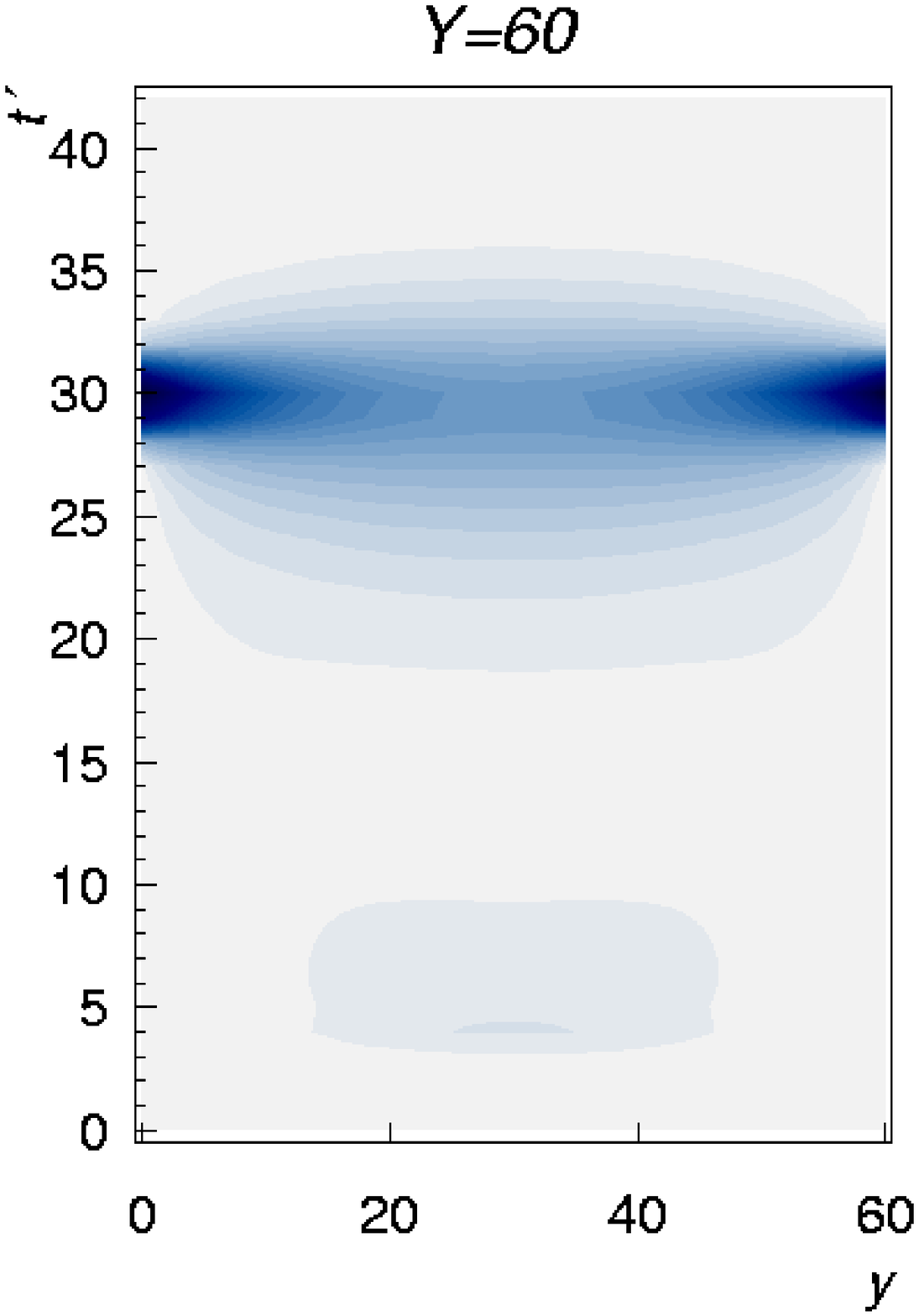,height=0.22\textheight}
    \epsfig{file=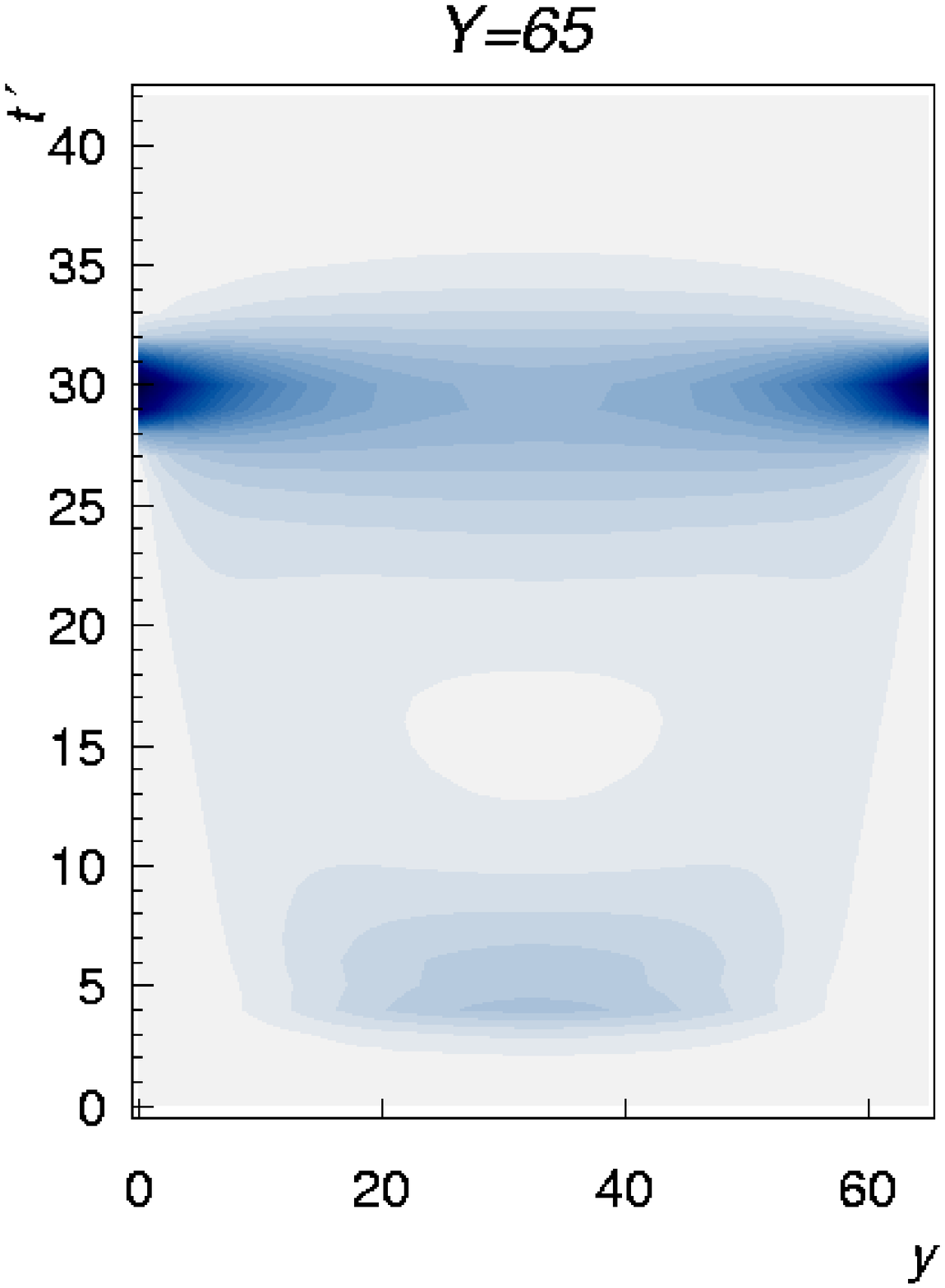,height=0.22\textheight}
    \vspace{0.2cm}\\
    \epsfig{file=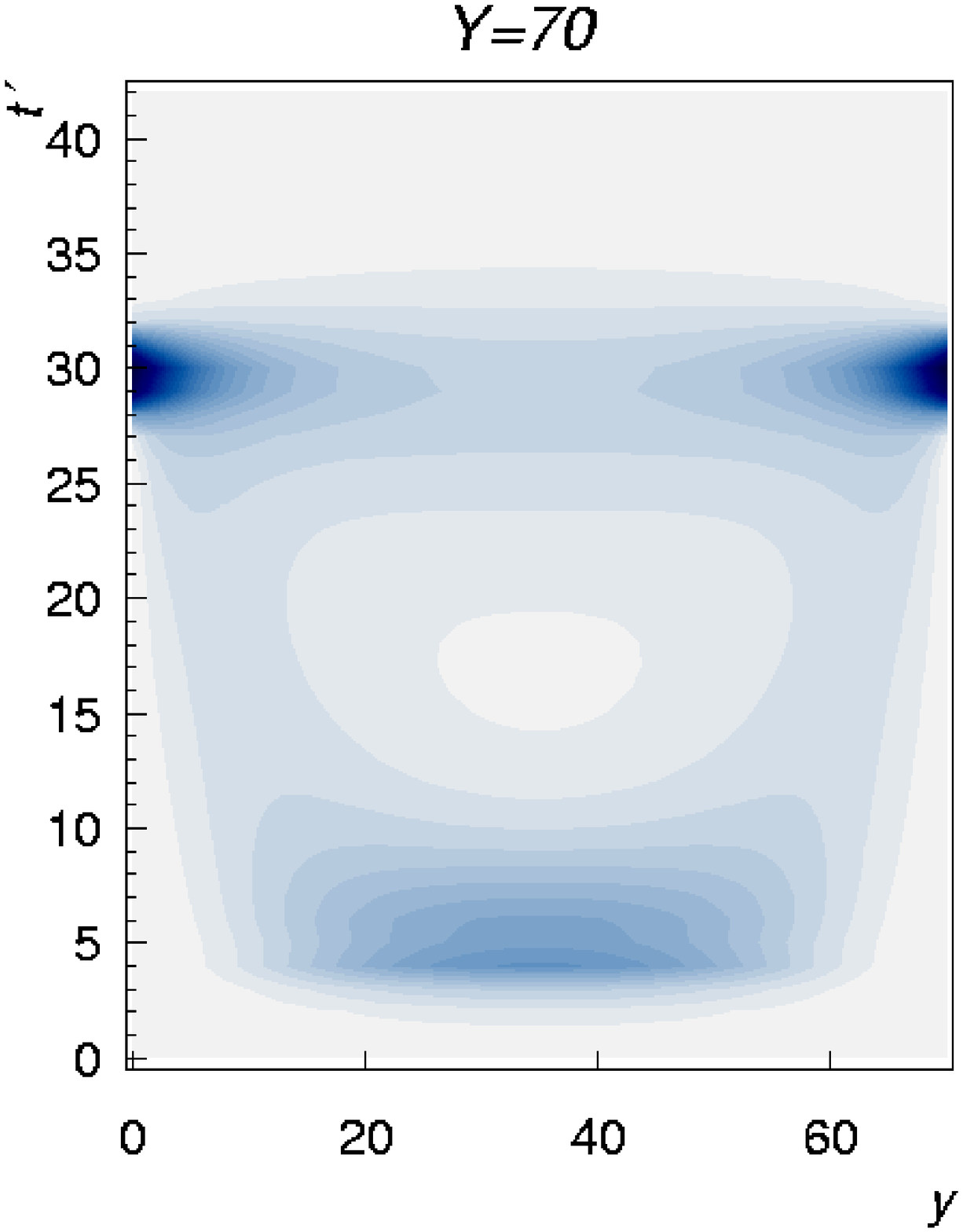,height=0.22\textheight}
    \epsfig{file=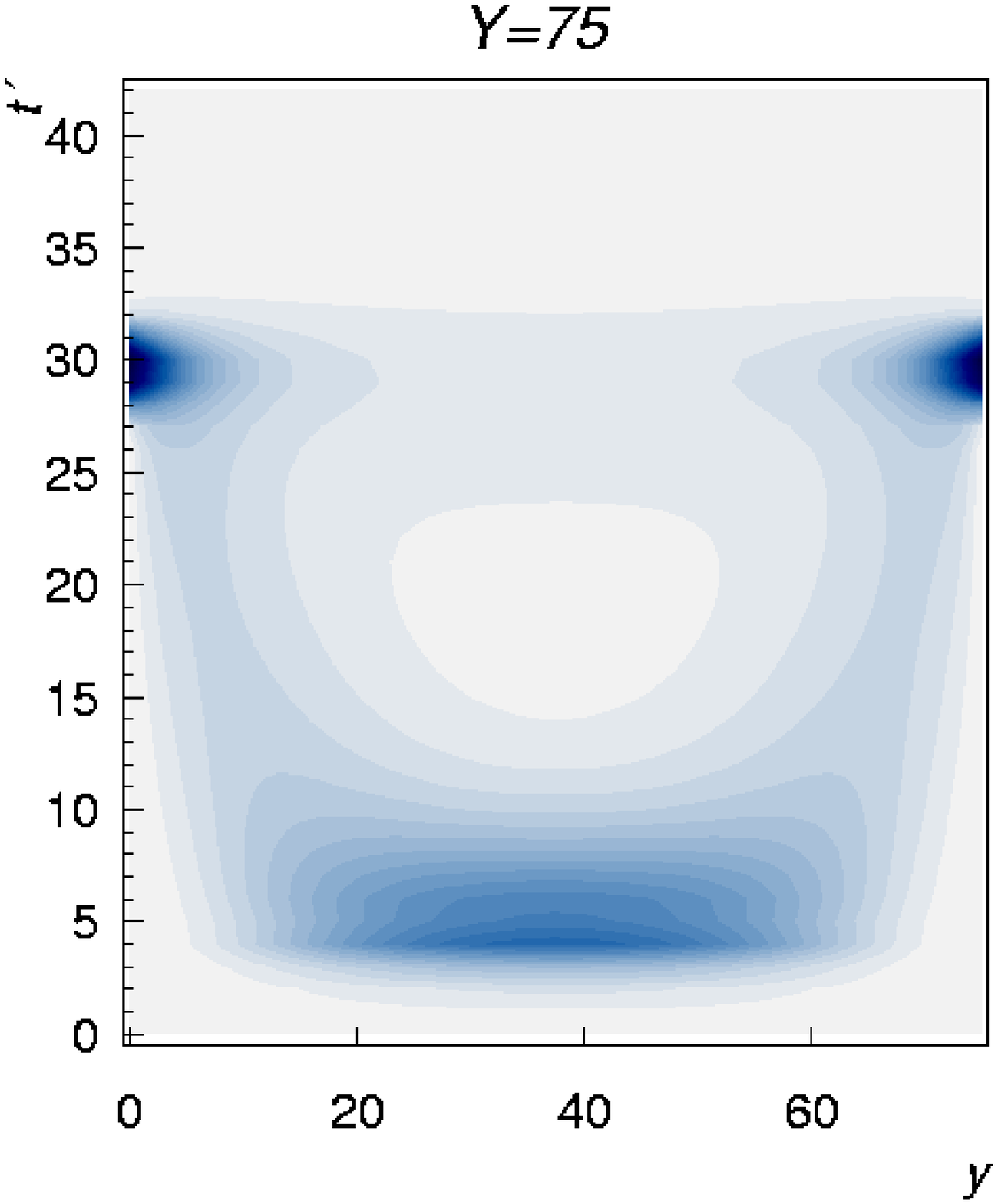,height=0.22\textheight}
    \epsfig{file=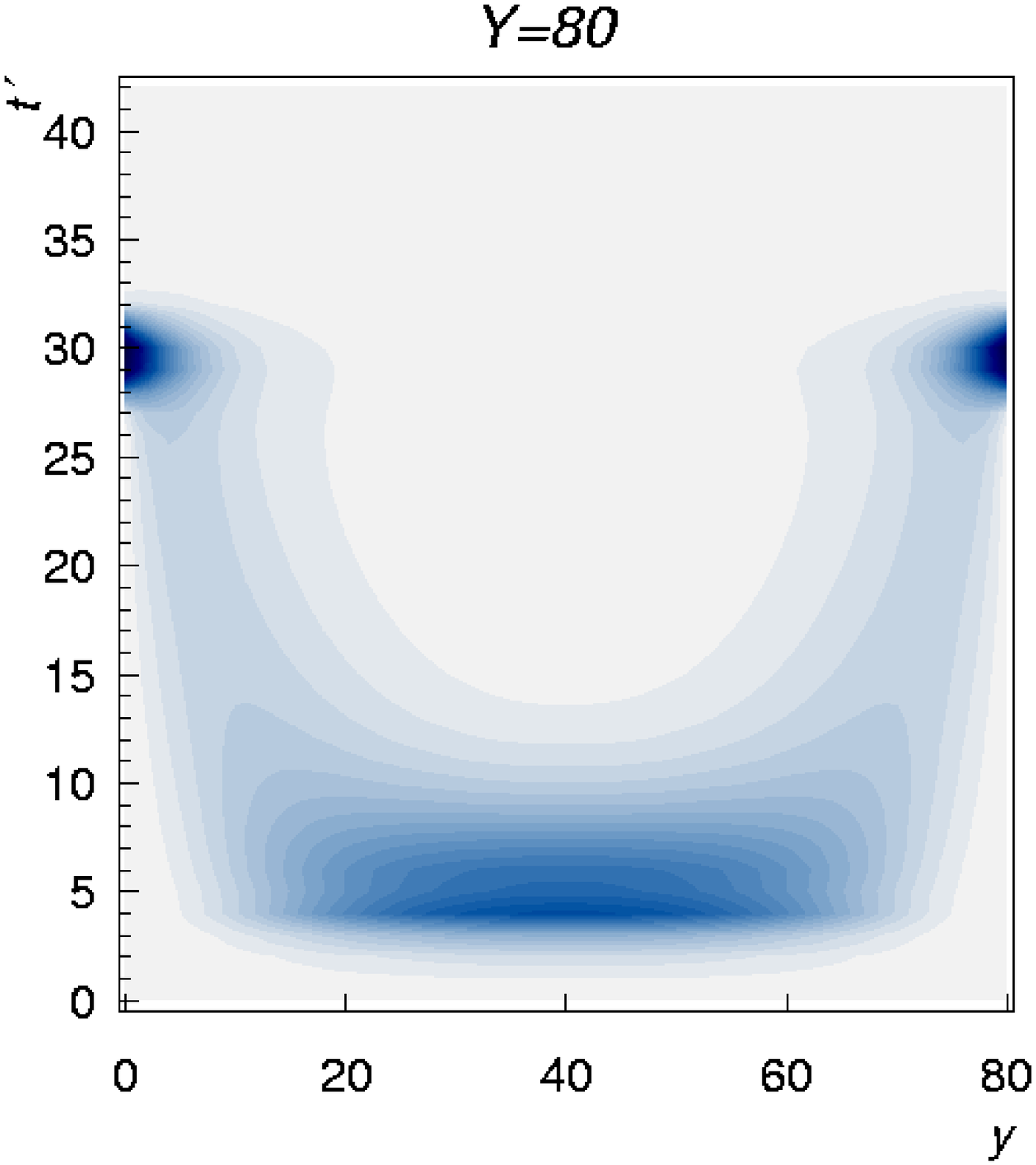,height=0.22\textheight}
    \vspace{0.2cm}\\
    \epsfig{file=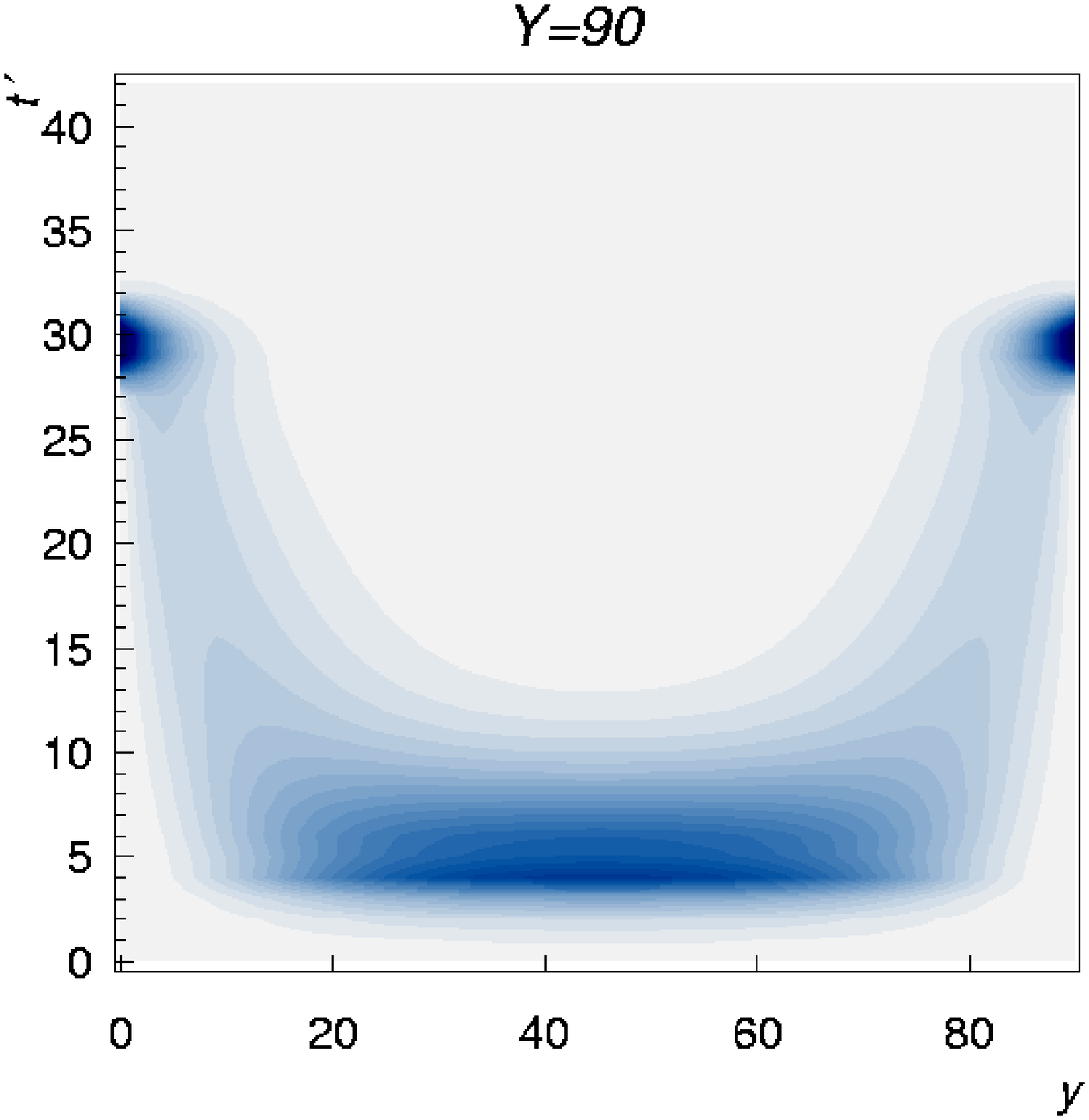,height=0.22\textheight}
    \epsfig{file=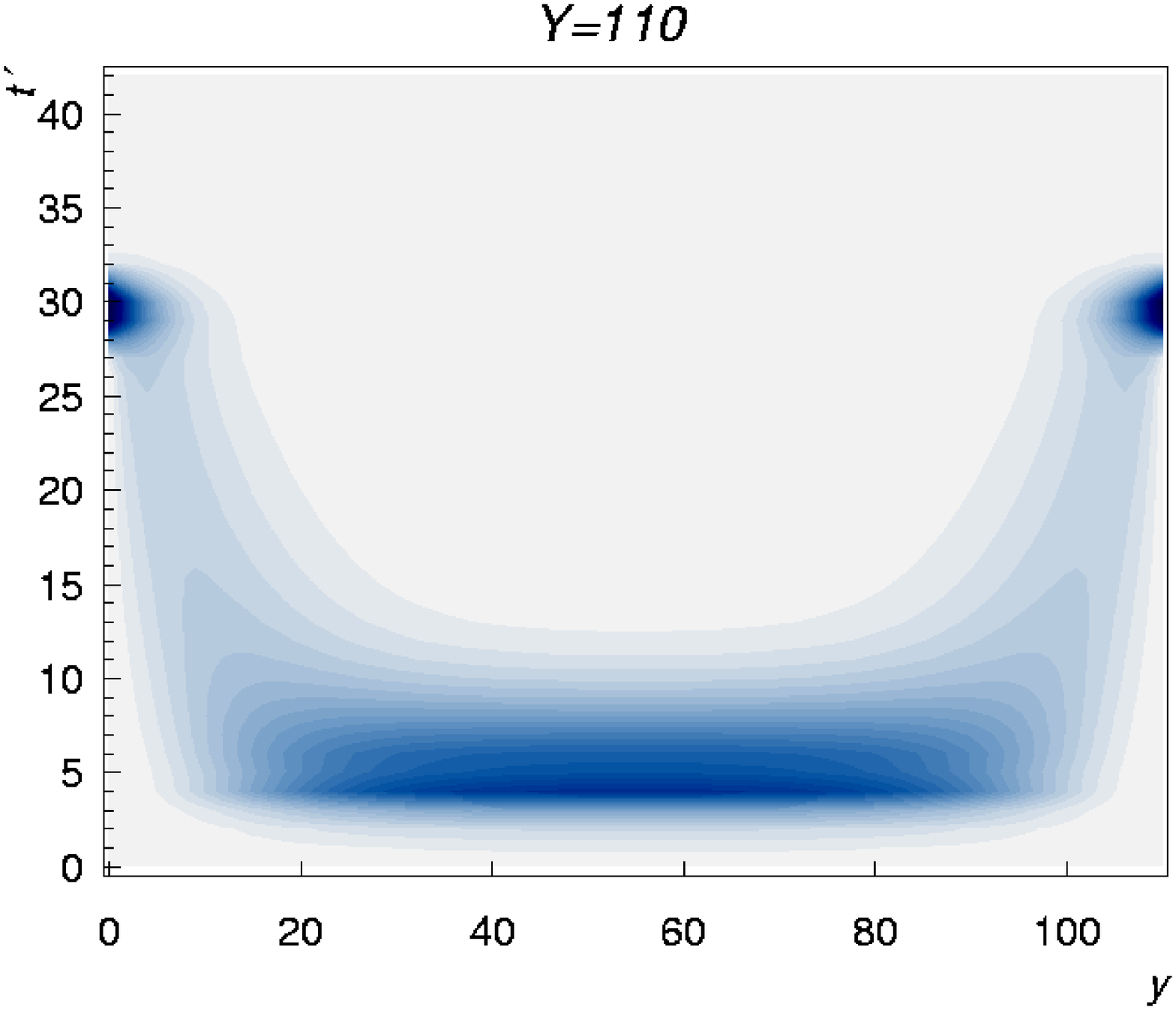,height=0.22\textheight}
    \caption{Contour plots for $f(Y,y,t,t')$, illustrating different
      stages of the evolution: $Y=30$ and $50$ illustrates standard
      `cigar' type plots, $Y=60$--$70$ show the point where tunneling
      begins to play a role, while for $Y \ge 80$
      the momentum configuration is only in the  non-perturbative regime.}
    \label{fig:cigars}
  \end{center}
\end{figure}

Tunneling transition will take place when $G_{\rm tunnel} \simeq G_{\rm pert}$
and this occurs at rapidity 
\begin{equation}
Y_{\rm tunnel}(t) = \frac{t-\bar{t}}{\omp - \oms(t)}
\label{eq:ytunnel}
\end{equation}
which shows the linear dependence on $t$ and the slope governed by the value of the intercept of the non-perturbative Pomeron $\omp$.

From this analysis it is clear that in the case of the small $x$ evolution with running coupling one has always 
contamination from  the non-perturbative contribution. The real problem
lies then in the definition of the perturbative hard Pomeron. 
A way of solving this problem is to consider the $b$-expansion \cite{BEXP} i.e. take the limit in which $b \rightarrow 0$
where $b$ is the beta function coefficient of the QCD coupling.
It has been shown \cite{BEXP} that the gluon Green's function can be decomposed into the perturbative and non-perturbative terms and   that the ratio of these two components is roughly of the form
\begin{equation}
\frac{G_{I\!\!P}(Y;t,t_0)}{G_{\rm pert}(Y;t,t_0)} \sim \exp{[(\omp-\oms(t)) Y-\frac{1}{b\asb(t)}g(\asb(t))]}
\end{equation}
where $g(\asb(t))$ is some function which can depend on the details of the model for small $x$ evolution (for example Airy, collinear, full BFKL).
We see, that the non-perturbative Pomeron is asymptotically leading since
we always have $\omp > \oms(t)$ but is suppressed by the universal exponential factor. By taking the limit $b\rightarrow 0$ we can eliminate the non-perturbative Pomeron and are able to isolate purely perturbative contribution
which is independent of the given regularisation procedure for the running coupling. 
The $b \rightarrow 0$ limit corresponds to the assumption of the very  slowly varying QCD coupling (expansion around the fixed coupling limit). Using the $b$ - expansion, see \cite{BEXP}, one is able to identify
systematically various types of diffusion corrections  $\sim b^2 \asb^5 Y^3, b^2 \asb^4 Y^2$ and specify the radius of convergence for this series
given by the parameter $\zeta_c=b\chi_m \bar{\asb}^2(t_0) \simeq 0.264$, where $\chi_m=4\ln 2$ being the minimum of the BFKL kernel eigenvalue.
We have also checked numerically that the perturbative behaviour indeed breaks down  at rapidities $Y\sim t^2$ which  can be seen from the above quoted form of the parameter $\zeta_c$.
In Fig.2 we show the maximum rapidity for which the perturbative part can be defined using two methods. In the first one, we calculate the solution to the equation (\ref{eq:bfkl}) with two different regularisations for $\asb(t)$ and define $Y_{\rm max}$ as the limiting rapidity at which the two solutions start to diverge - dashed line.
In the second method we apply the $b$-expansion method, and calculate the perturbative gluon Green's function from the following series $\ln G_{\rm pert}(Y;t,t)= \sum_i b^i l_i(\asb(t),Y)$. We then truncate this series at different orders and define again $Y_{\rm max}$ as the divergence point.
We clearly see from Fig.2 that the first method yields linear dependence on $t$ which is consistent with formula (\ref{eq:ytunnel}) and suggest the tunneling as the relevant mechanism for the breakdown of perturbative evaluation. In the second case, we observe $\sim t^2$ behaviour which is consistent with the prediction based on the critical value $\zeta_c$. Thus the perturbative prediction 
can be in principle made up to the values $Y_{\rm max} \sim t^2$.

\begin{figure}[htbp]
  \begin{center}
    \epsfig{file=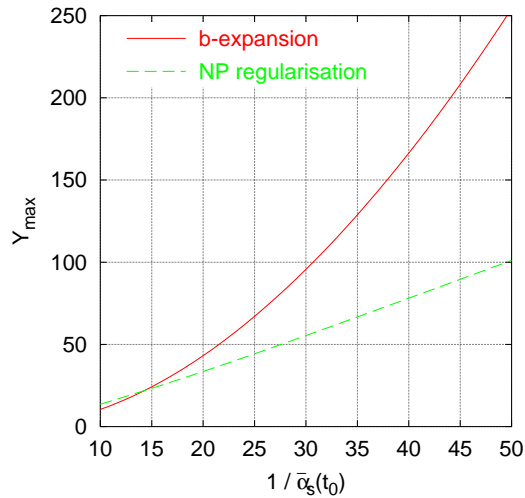,width=0.54\textwidth}
    \caption{The maximum perturbatively accessible value of $Y$, as a
      function of $1/\asb(t_0)$, determined by comparing different
      non-perturbative regularisations of $\alpha_s$, or different
      truncations of the $b$-expansion. }
    \label{fig:ymax}
  \end{center}
\end{figure}

\section{Acknowledgments}
The results presented in this talk have been obtained in collaboration with
M. Ciafaloni, D. Colferai and G.P. Salam.
Work supported in part by the E.U. QCDNET
 contract FMRX-CT98-0194 and   by the Polish  
Committee for Scientific Research (KBN) grants no. 2P03B 05119, 2P03B
12019, 5P03B 14420.

\end{document}